\documentclass[twocolumn,showpacs,aps,groupedaddress,prd,eqsecnum]{revtex4}
\usepackage{amsfonts,latexsym,eucal}
\usepackage[dvips]{graphicx}

\def\beq{\begin{equation}}
\def\eeq{\end{equation}}
\def\bea{\begin{eqnarray}}
\def\eea{\end{eqnarray}}

\def\pa{\partial}
\def\ra{\rightarrow}
\def\mn{_{\mu\nu}}

\def\tt{\tilde{t}}
\def\tr{\tilde{r}}
\def\tp{\tilde{\phi}}
\def\tpp{{\tp_{~}}'}
\def\tpps{{\tp_{~}}^{\prime2}}
\def\tv{\tilde{V}}

\def\tm{\tilde{m}}

\def\to{\tilde{\omega}}
\def\bp{\mbox{\boldmath$\phi$}}

\def\tE{\tilde{E}}
\def\tQ{\tilde{Q}}
\newcommand{\dalm}{\kern1pt\vbox{\hrule height 0.9pt\hbox{\vrule width 0.9pt
\hskip 2.5pt\vbox{\vskip 5.5pt}\hskip 3pt\vrule width 0.3pt}\hrule height 0.3pt}\kern1pt}

\usepackage{epsfig}

\begin{document}

\thispagestyle{empty}

\title{How does gravity save or kill Q-balls?}
\author{Takashi Tamaki}
\email{tamaki@ge.ce.nihon-u.ac.jp}
\affiliation{Department of Physics, General Education, College of Engineering, 
Nihon University, Tokusada, Tamura, Koriyama, Fukushima 963-8642, Japan}
\author{Nobuyuki Sakai}
\email{nsakai@e.yamagata-u.ac.jp}
\affiliation{Department of Education, Yamagata University, Yamagata 990-8560, Japan}
\date{\today}

\begin{abstract}
We explore stability of gravitating Q-balls with potential
$V_4(\phi)={m^2\over2}\phi^2-\lambda\phi^4+\frac{\phi^6}{M^2}$
via catastrophe theory, as an extension of our previous work on Q-balls with potential
$V_3(\phi)={m^2\over2}\phi^2-\mu\phi^3+\lambda\phi^4$.
In flat spacetime Q-balls with $V_4$ in the thick-wall limit are 
unstable and there is a minimum charge $Q_{{\rm min}}$, 
where Q-balls with $Q<Q_{{\rm min}}$ are nonexistent.
If we take self-gravity into account, on the other hand, there exist 
stable Q-balls with arbitrarily small charge, no matter how weak 
gravity is. That is, gravity saves Q-balls with small charge. 
We also show how stability of Q-balls changes as gravity becomes strong. 

\end{abstract}

\maketitle

\section{Introduction}

Q-balls \cite{Col85}, a kind of non-topological solitons \cite{LP92}, appear in a large family of field
theories with global U(1) (or more) symmetry, and could play important roles in
cosmology. For example, the Minimal Supersymmetric Standard Model may contain
baryonic Q-balls, which could be responsible for baryon asymmetry \cite{SUSY} and dark
matter \cite{SUSY-DM}.

Because Q-balls are typically supposed to be microscopic objects, their self-gravity is usually ignored.
Therefore, stability of Q-balls has been intensively studied in flat spacetime  \cite{stability,PCS01,SakaiSasaki}.
Q-balls in arbitrary dimension \cite{Copeland} and spinning Q-balls \cite{Volkov, KKL05} have also been studied.

If  Q-balls are so large or so massive, on the other hand, their size becomes astronomical and their gravitational effects are remarkable \cite{Grav-Q,KKL05}.
For example, it has been shown \cite{multamaki} that the size of Q-balls is bounded above due to gravity.
There are analogous objects which are analogous to gravitating Q-balls: boson stars  \cite{boson-review}.
While Q-balls exist even in flat spacetime, boson stars are supported by gravity and nonexistent in flat spacetime.
Although a difference in theory between Q-balls and boson stars is solely the potential parameters, investigations of their properties have been carried out separately so far.

In our previous paper \cite{TamakiSakai}, to obtain a unified picture of Q-balls and boson stars, we made an analysis of gravitating Q-balls and boson stars via catastrophe theory~\cite{PS78}.
In Ref.\cite{TamakiSakai} we chose a potential for Q-balls
\beq\label{V3}
V_{3}(\phi):={m^2\over2}\phi^2-\mu\phi^3+\lambda\phi^4,
~~~{\rm with} ~~~ m^2,~\mu,~\lambda>0,
\eeq
because in the limit of $\mu\ra0$ this approaches a typical potential for boson stars,
\beq\label{VBS}
V_{BS}(\phi):={m^2\over2}\phi^2+\lambda\phi^4,
~~~{\rm with} ~~~ m^2,~\lambda>0.
\eeq
As a result, we found that Q-balls and boson stars expose a similar phase 
relation between a charge and a total Hamiltonian energy. 
(See, cusp structures in Figs.1(a) and 12(a) in \cite{TamakiSakai}.)

In this paper we extend our analysis via catastrophe theory to a potential
\beq\label{V4}
V_4(\phi):={m^2\over2}\phi^2-\lambda\phi^4+\frac{\phi^6}{M^2} 
~~~{\rm with} ~~~ m^2,~\lambda,~M>0,
\eeq
which we call $V_4$ Model~\cite{foot1}.
We choose this potential because previous work on Q-balls in flat spacetime \cite{PCS01,SakaiSasaki} showed stability of Q-balls with $V_3$ Model (\ref{V3}) and $V_4$ Model (\ref{V4}) are quite different.
We are interested in how gravitating Q-balls properties depend on potentials and what universal properties are.

This paper is organized as follows.
In Sec. II, we derive equilibrium field equations. 
In Sec. III, we show numerical results of equilibrium Q-balls and discuss their stability. 
In Sec. IV, we discuss why thick-wall solutions become stable against the naive expectation that 
gravity is not effective for Q-balls with small charge. 
In Sec. V, we devote to concluding remarks.

\section{Analysis method of equilibrium Q-balls}

\subsection{Equilibrium field equations}

We begin with the action
\beq\label{Sg}
{\cal S}=\int d^4x\sqrt{-g}\left\{ \frac{{\cal R}}{16\pi G}-\frac12g^{\mu\nu}\pa_{\mu}\bp\cdot\pa_{\nu}\bp
-V(\phi)\right\}, 
\eeq
where $\bp=(\phi_1,~\phi_2)$ is an SO(2)-symmetric scalar field and 
$\phi:=\sqrt{\bp\cdot\bp}=\sqrt{\phi_1^2+\phi_2^2}$.
We assume a spherically symmetric and static spacetime, 
\beq\label{metric1}
ds^2=-\alpha^2(r)dt^2+A^2(r)dr^2+r^2(d\theta^2+\sin^2\theta d\varphi^2).
\eeq

For the scalar field, we assume that it has a spherically symmetric and stationary form, 
\beq\label{phase}
(\phi_1,\phi_2)=\phi(r)(\cos\omega t,\sin\omega t).
\eeq
Then the field equations become
\bea\label{Gtt}
-{r A^3\over2}G^t_t&:=&A'+{A\over2r}(A^2-1) \nonumber \\
&=&{4\pi G}r A^3\left({{\phi'}^2\over2A^2}
+{\omega^2\phi^2\over2\alpha^2}+V\right),
\\\label{Grr}
{r\alpha\over2}G_{rr}&:=&\alpha'+{\alpha\over2r}(1-A^2) \nonumber \\
&=&{4\pi G}r\alpha A^2
\left({{\phi'}^2\over2A^2}+{\omega^2\phi^2\over2\alpha^2}-V\right),
\\\label{Box}
{A^2\phi\over\phi_1}\Box\phi_1&:=&
\phi''+\left(\frac2r+{\alpha'\over\alpha}-{A'\over A}\right)\phi'
+\left({\omega A\over\alpha}\right)^2\phi \nonumber \\
&=&A^2{dV\over d\phi},
\eea
where $':= d/dr$. 
To obtain Q-ball solutions in curved spacetime, we should solve 
(\ref{Gtt})-(\ref{Box}) with boundary conditions, 
\bea
&& A(0)=A(\infty)=\alpha(\infty)=1,\nonumber \\ 
&& A'(0)=\alpha'(0)=\phi'(0)=\phi(\infty)=0.
\label{bcg}
\eea
We also restrict our solutions to monotonically decreasing $\phi (r)$. 
Due to the symmetry, there is a conserved charge called Q-ball charge,
\bea\label{Q}
Q&:= &\int d^3x\sqrt{-g}g^{0\nu}(\phi_1\pa_\nu\phi_2-\phi_2\pa_\nu\phi_1)=\omega I,
\nonumber  \\
&&{\rm where}~~~
I:=4\pi\int{A r^2\phi^2\over\alpha}dr.
\eea

We suppose $V_4$ Model (\ref{V4}).
Rescaling the quantities as
\bea
&&\tt :=\lambda Mt,~~ \tr :=\lambda Mr,~~
\tp :={\phi\over\sqrt{\lambda}M},~~\nonumber  \\
&&\tv_4 :={V_4\over\lambda^3M^4}=\frac{\tm^2}{2}\tp^2 -\tp^4 +\tp^6,~~ \nonumber  \\
&&\tm :={m\over\lambda M},~~
\to :={\omega\over\lambda M},~~\kappa := G\lambda M^2, 
\label{rescale}
\eea
the field equations (\ref{Gtt})-(\ref{Box}) with the potential 
(\ref{V4}) are rewritten as
\beq\label{rsfe1}
A'+{A\over2\tr}(A^2-1)
=4\pi\kappa\tr A^3\left({\tpps\over2A^2}+{\to^2\tp^2\over2\alpha^2}+\tv_{4}\right),
\eeq\beq\label{rsfe2}
\alpha'+{\alpha\over2\tr}(1-A^2)
=4\pi\kappa\tr\alpha A^2\left({\tpps\over2A^2}+{\to^2\tp^2\over2\alpha^2}-\tv_{4}\right),
\eeq\beq\label{rsfe3}
\tp^{\prime\prime}+\left(\frac2{\tr}+{\alpha'\over\alpha}-{A'\over A}\right)\tpp
+\left({\to A\over\alpha}\right)^2\tp=A^2{d\tv_{4}\over d\tp}.
\eeq

\subsection{Stability analysis method via catastrophe theory}

In our previous paper~\cite{TamakiSakai}, we discussed how we apply catastrophe theory to the Q-ball and boson star systems. 
Here, we summarize our method. An essential point is to choose {\it behavior variable}({\it s}),
{\it control parameter}({\it s}) and a {\it potential\/} in the Q-ball system appropriately. 

We use the Hamiltonian energy $E$ as a {\it potential\/} because $\delta E/\delta\phi|_Q = \delta E/\delta g\mn=0$, 
reproduces the equilibrium field equations (\ref{Gtt})-(\ref{Box}).
The Hamiltonian energy $E$ was calculated as \cite{TamakiSakai}
\beq\label{H}
E=\lim_{r\ra\infty}{r^2\alpha'\over2GA}={M_S\over2},
\eeq
where $M_S$ is the Schwarzschild mass.
We also normalize $E$ and $Q$ as
\beq\label{Enormalize}
\tE := {E\over M},~~ \tQ :=Q\lambda.
\eeq

Because the charge $\tQ$ and the model parameters $\tilde{m}^2$ and $\kappa$ 
specify the system environment, they should be regarded as {\it control parameters}. 
To discuss a {\it behavior variable} we consider a one-parameter family of perturbed field configurations $\phi_{x}(r)$ near the equilibrium solution $\phi(r)$.
Because $dE[\phi_x]/dx=(\delta E/\delta\phi_x)d\phi_x/dx= 0$ when $\phi_{x}$  is an equilibrium solution, $x$ is  a {\it behavior variable}. 

According to Thom's theorem, if the system has two control parameters, 
there is essentially one behavior variable;  if the system has three control parameters, there are one or two behavior variables. Because the present Q-ball system 
contains $(\tilde{Q}, \tilde{m}^2, \kappa)$, we speculate that each has two {\it behavior variables}, 
$\tilde{\omega}^{2}$ and $\tilde{\phi}(0)$.
However, because stability structure of equilibrium solutions in three-parameter space 
$(\tilde{Q}, \tilde{m}^2, \kappa)$ 
is very complicated and our interest is how gravitational effects change the stability structure, in what follows, we discuss the stability structure of equilibrium solutions in two-parameter space $(\tilde{Q}, \kappa)$ 
under fixed $\tilde{m}^2$. 

Our method of analyzing the stability of Q-balls is as follows.
\begin{itemize}
\item Fix the value of $\tilde{m}^2$.
\item
Solve the field equations~(\ref{Gtt})-(\ref{Box}) with the boundary 
condition~(\ref{bcg}) numerically to obtain equilibrium 
solutions $\tilde{\phi}(r)$ for various values of $\tilde{\omega}$ and $\kappa$.
\item 
Calculate $\tilde{Q}$ for each solution to obtain the {\it equilibrium space} 
${\cal M}=\{(x,\tilde{Q},\kappa)\}$.
We denote the equation that determines ${\cal M}$ by $f(x,\tilde{Q},\kappa)=0$.
\item 
Find folding points where $\pa \tilde{Q}/\pa x=0$ or $\pa\kappa/\pa x=0$, 
in ${\cal M}$, which are identical to the stability-change points, 
$\Sigma=\{(x,\tilde{Q},\kappa)\,|\,{\pa f/\pa x}=0, ~f=0\}$.
\item
Calculate the energy $\tilde{E}$ by (\ref{H}) for equilibrium solutions
around a certain point in $\Sigma$ to find whether the point is 
a local maximum or a local minimum. Then we find the stability 
structure for the whole ${\cal M}$.
\end{itemize}

\section{Equilibrium solutions and their stability}

In preparation for discussing gravitating Q-balls, we review their equilibrium solutions and stability in flat spacetime ($\kappa=0$).
The scalar field equation (\ref{rsfe3}) reduces to
\beq\label{rsfeflat}
\tp^{\prime\prime}=-\frac{2}{\tilde{r}}\tpp-\tilde{\omega}^2\tilde{\phi}+{d\tilde{V}_{4}\over d\tilde{\phi}}\,.
\eeq
This is equivalent to the field equation for a single static scalar 
field with the potential $V_{\omega}:= \tilde{V}_{4}-\tilde{\omega}^2\tilde{\phi}^2/2$.
Equilibrium solutions satisfying boundary conditions (\ref{bcg}) 
exist if min$(V_{\omega})<\tilde{V}_{4}(0)$ and $d^2V_{\omega}/d\tilde{\phi}^2(0)>0$,
which is equivalent to
\beq\label{omega}
0<\epsilon^2 <\frac{1}{2}, 
\eeq
where $\epsilon:=\sqrt{\tilde{m}^{2}-\tilde{\omega}^2}$.
The two limits $\epsilon^2\ra\frac{1}{2}$ and $\epsilon\ra 0$ correspond to the thin-wall limit and the thick-wall limit, respectively.

It is usually assumed that the potential has an absolute minimum at $\phi=0$. If
$V(0)$ is a local minimum but the absolute minimum is located at $\phi\ne0$, true vacuum
bubbles with charge (Q-bubbles) may appear.
The condition for Q-bubbles is $\tilde{m}^2 < 0.5$.
Therefore, stability structure falls into two classes,  
$\tilde{m}^2 < 0.5$ and $\tilde{m}^2 \geq 0.5$ \cite{SakaiSasaki}:
\begin{itemize}
\item $\tilde{m}^2 \geq 0.5$: For each $\tilde{m}^2$, there is a nonzero minimum charge, 
$\tQ_{\rm min}$, below which equilibrium solutions do not exist. 
For $\tQ >\tQ_{\rm min}$, stable and unstable solutions coexist. 
\item $\tilde{m}^2 < 0.5$: For each $\tilde{m}^2$, there is a maximum charge, 
$\tQ_{\rm max}$, as well as a minimum charge, $\tQ_{\rm min}$, where one stable 
and two unstable solutions coexist for $\tQ_{\rm min}<\tQ <\tQ_{\rm max}$. 
For $\tQ<\tQ_{\rm min}$ or $\tQ >\tQ_{\rm max}$, there is one unstable solution. 
\end{itemize}

To discuss gravitational effects later, it is useful to estimate the central value $\tp (0)$ in flat spacetime.
Because $V_{\omega}=0$ at spacial infinity, its order of magnitude 
is estimated as a solution of $V_{\omega}=0$ ($\tp (0)\neq 0$). 
For $V_4$ with the thick-wall condition $\epsilon \ll 1$, we obtain 
\beq\label{boundary-value}
\tp^2 (0)\simeq\frac{1-\sqrt{1-2\epsilon^2}}{2}\simeq \frac{\epsilon^2}{2}. 
\eeq
Thus, $\tp (0)\sim \epsilon$. 

It was shown \cite{TamakiSakai} that in $V_3$ Model properties of gravitating Q-balls also depend 
on whether $\tm^2\ge0.5$ or $\tm^2<0.5$. In the following analysis, therefore, we 
choose $\tilde{m}^2 =0.6$ and $0.3$ typically. Other cases are not qualitatively different from these cases. 
For our numerical calculation, we use the Bulirsch-Stoer method based on 
the double precision FORTRAN program. 

\begin{figure}[htbp]
\psfig{file=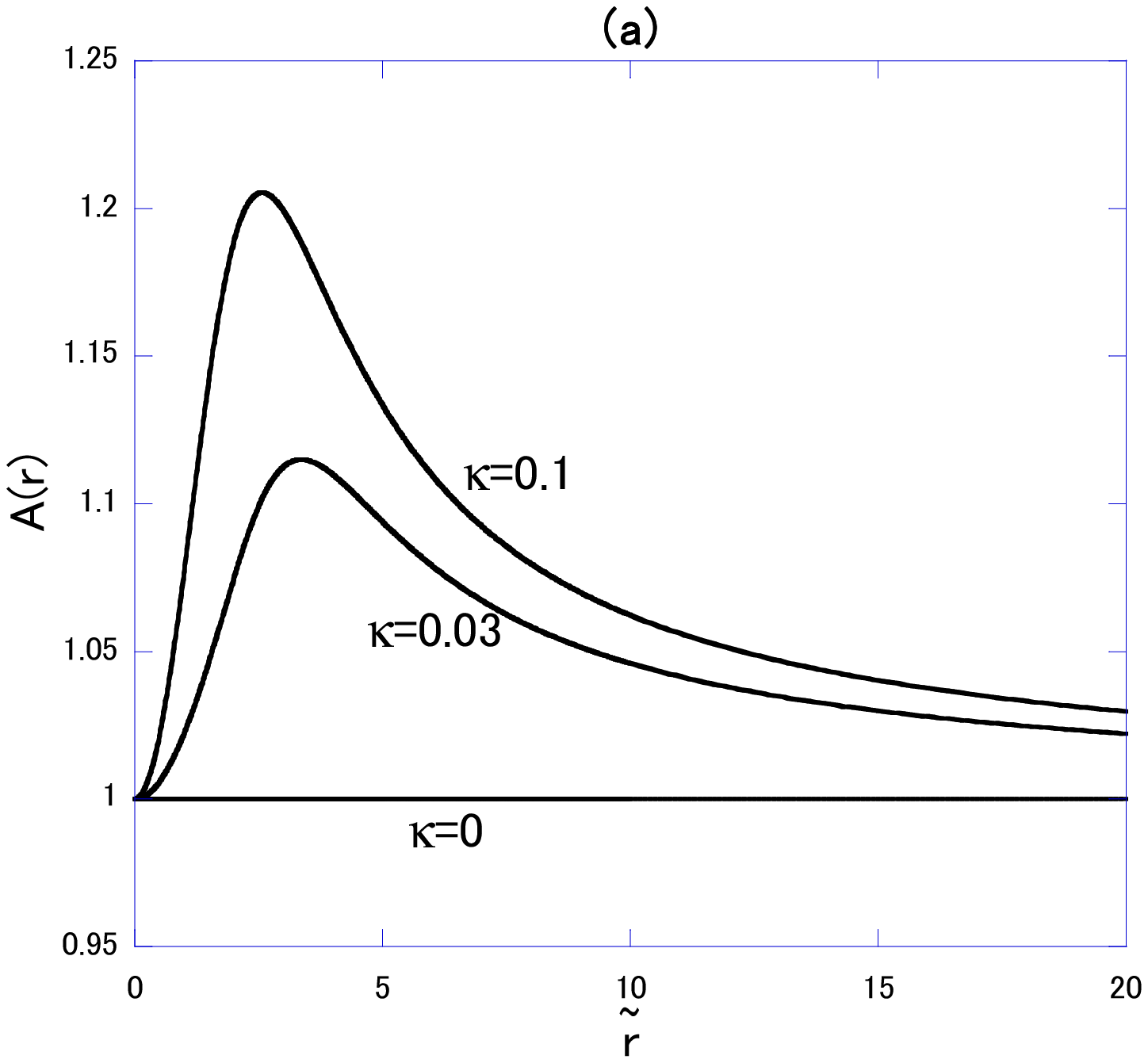,width=3in} \\
\psfig{file=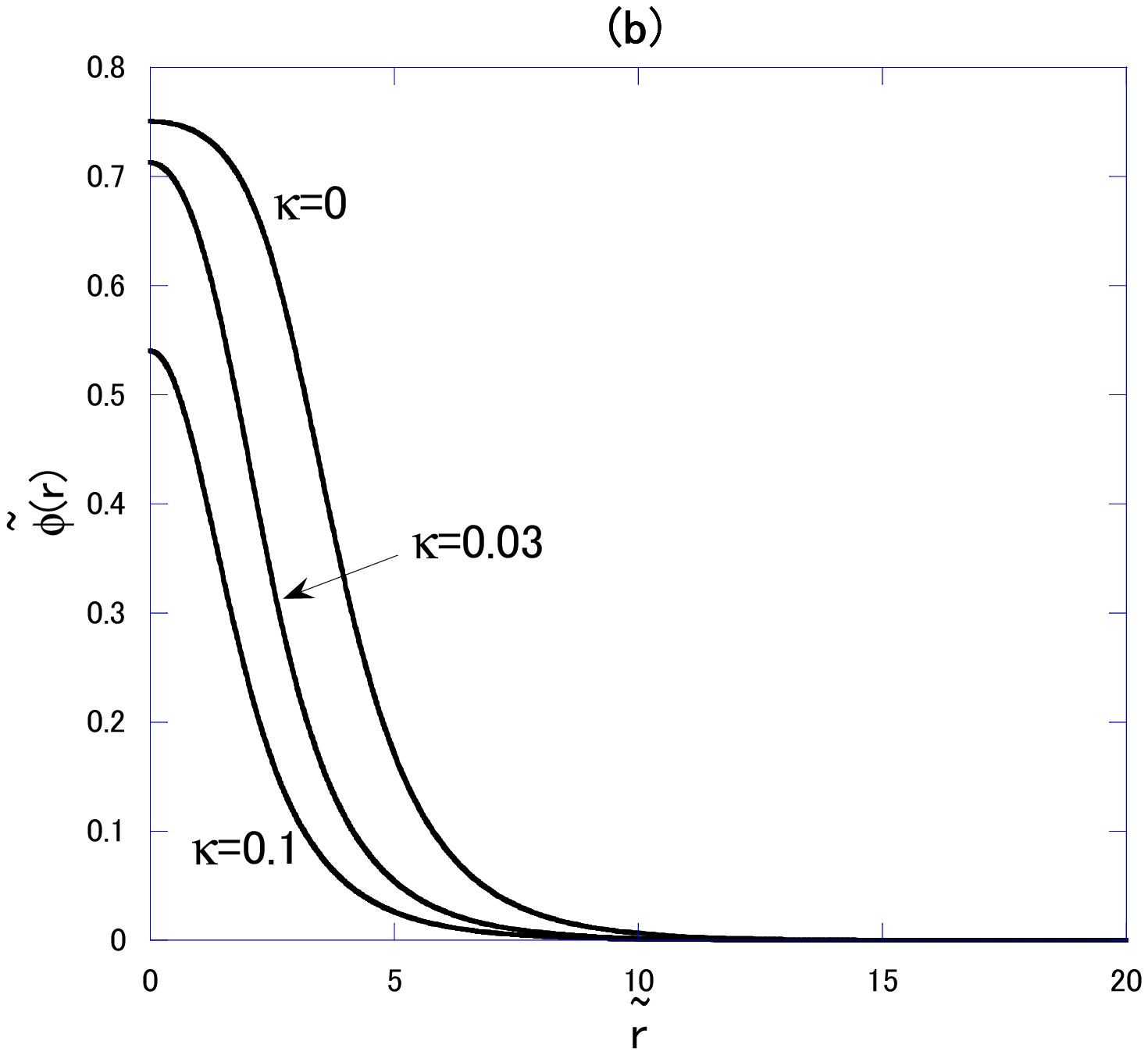,width=3in} 
\caption{Behavior of the metric $A(r)$ and $\tilde{\phi}(r)$ for the solutions 
$\tilde{\omega}^{2}\simeq 0.34$ in (a) and (b), respectively. 
We find that the scalar field is concentrated near the origin as shown in (b). 
This tendency becomes stronger as gravity is stronger. 
Thus, $A(r)$ varies near the origin compared with that in the thick-wall solutions. 
\label{m06thin} }
\end{figure}
\begin{figure}[htbp]
\psfig{file=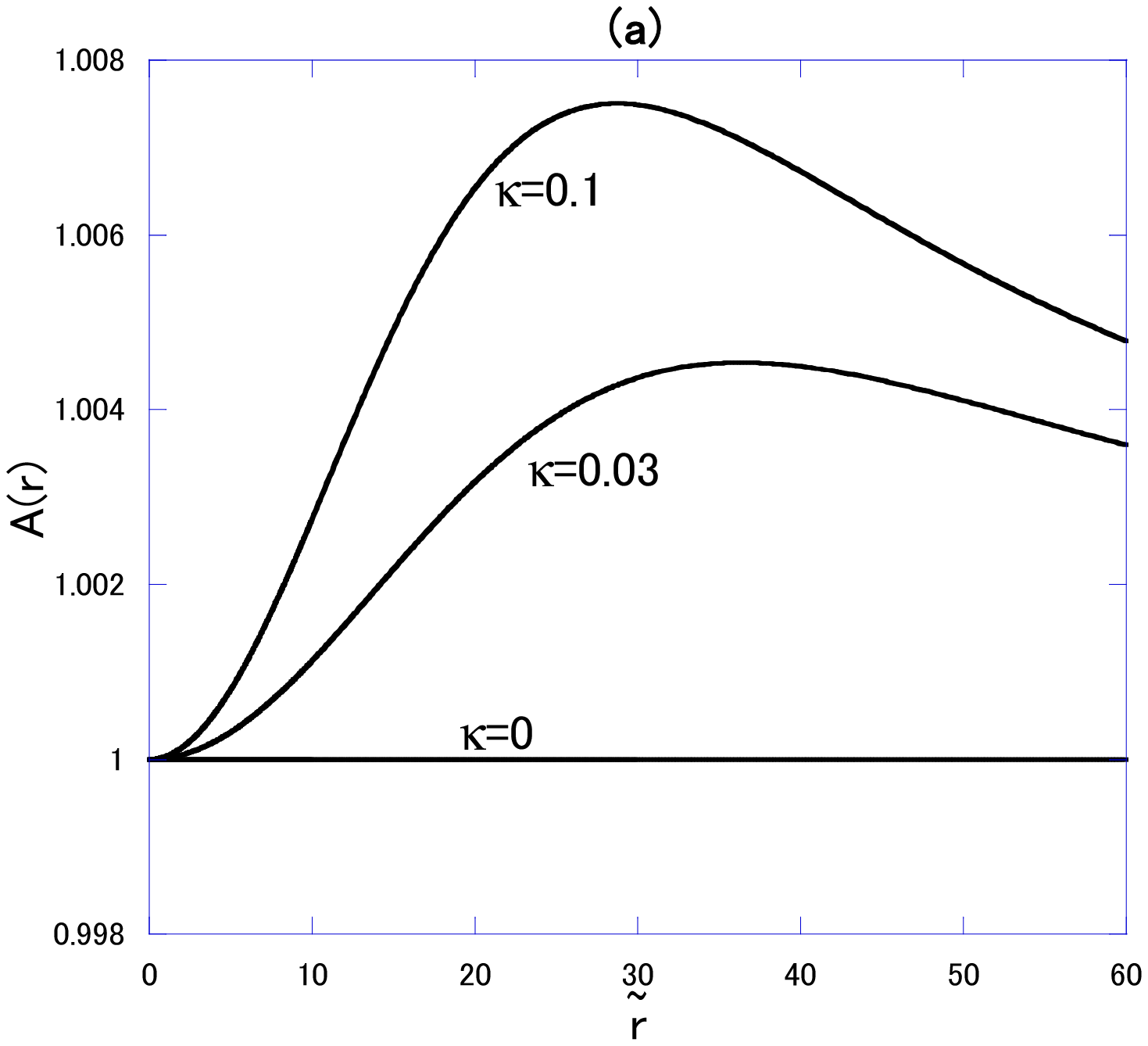,width=3in} \\
\psfig{file=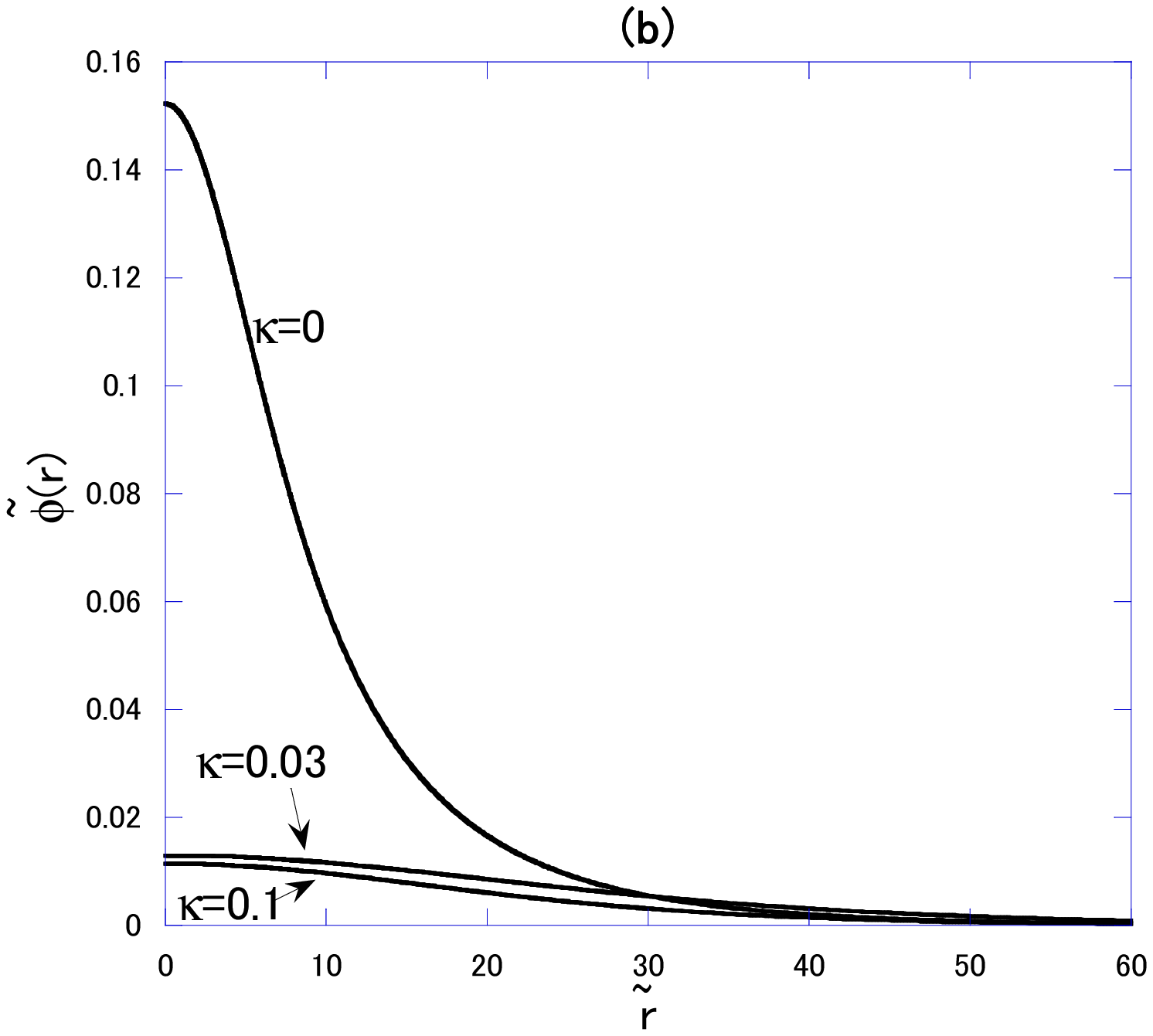,width=3in} 
\caption{Behavior of the metric $A(r):=\sqrt{g_{rr}}$ and $\tilde{\phi}(r)$ for the thick-wall solutions 
$\tilde{\omega}^{2}\simeq 0.595$ in (a) and (b), respectively. 
\label{m06thick} }
\end{figure}

\begin{figure}[htbp]
\psfig{file=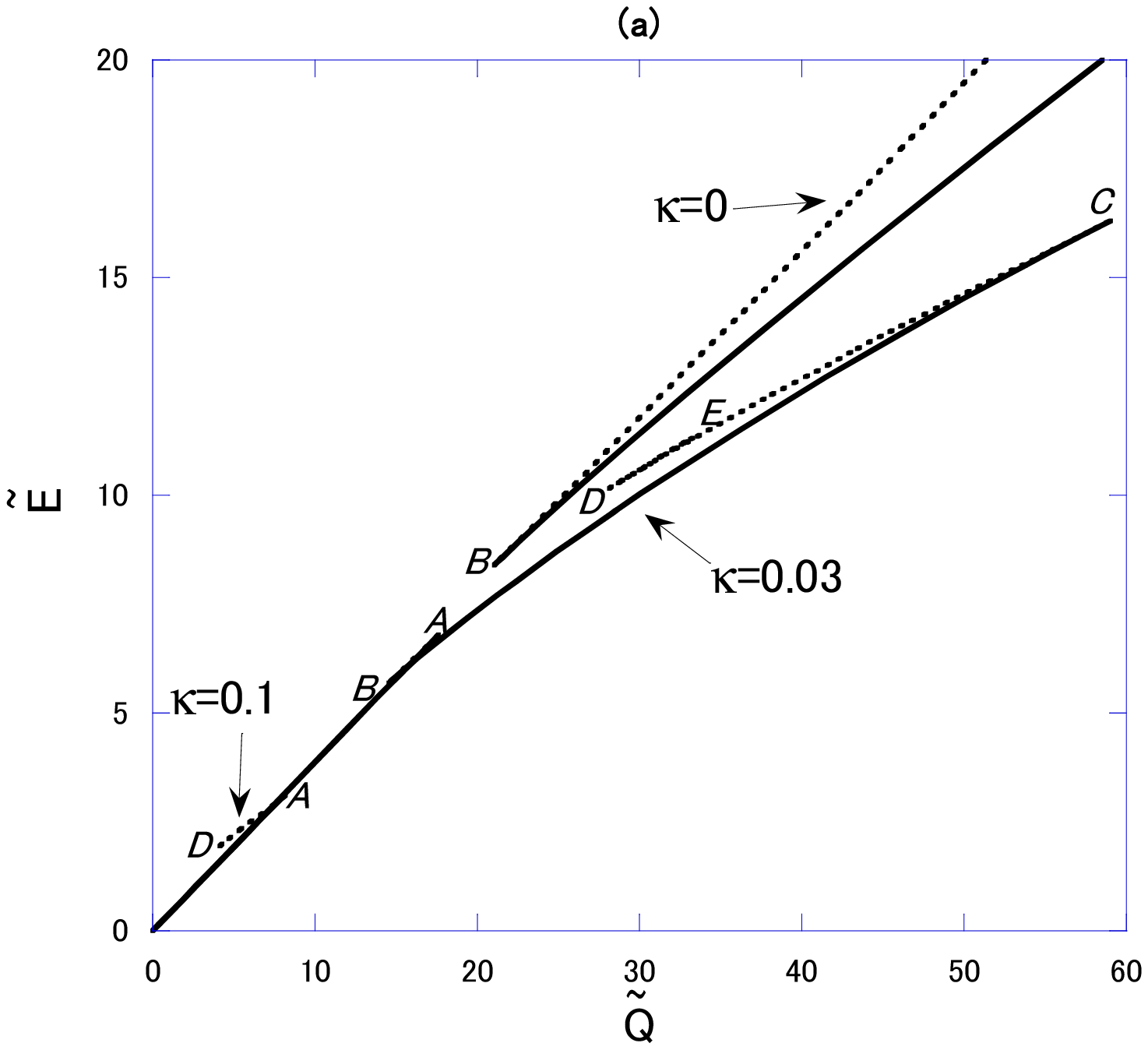,width=3in}  \\
\psfig{file=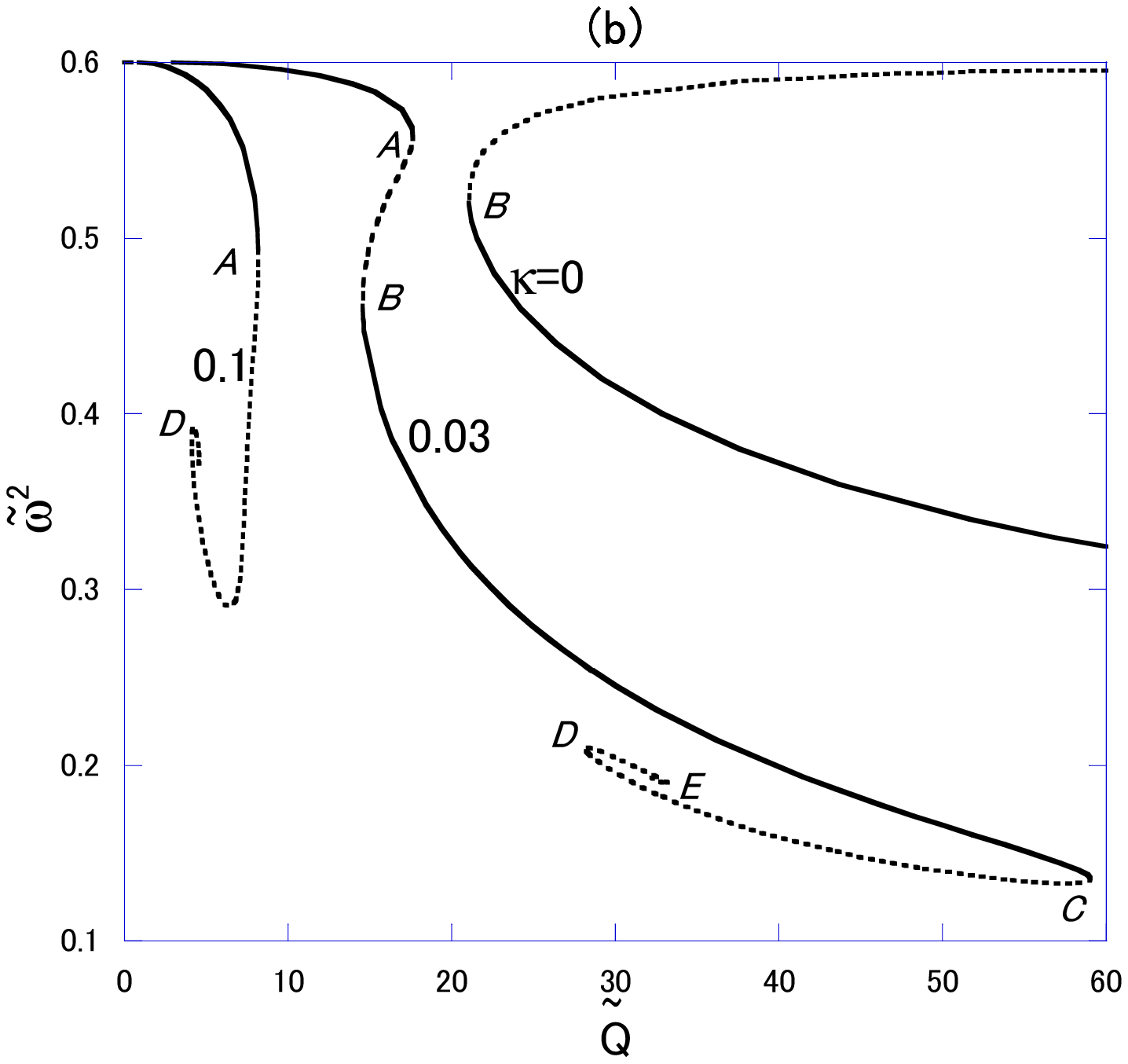,width=3in}
\caption{(a) $\tilde{Q}$-$\tilde{E}$ and (b) $\tilde{Q}$-$\tilde{\omega}^2$ relations for 
$\tilde{m}^{2}=0.6$ with $\kappa =0$, $0.03$ and $0.1$. For the flat case $\kappa =0$, it has 
been found that solutions with solid (dotted) lines are stable (unstable)~\cite{SakaiSasaki}. 
We extend these interpretations for the gravitating case. 
\label{QEml05}}
\end{figure}
\begin{figure}[htbp]
\psfig{file=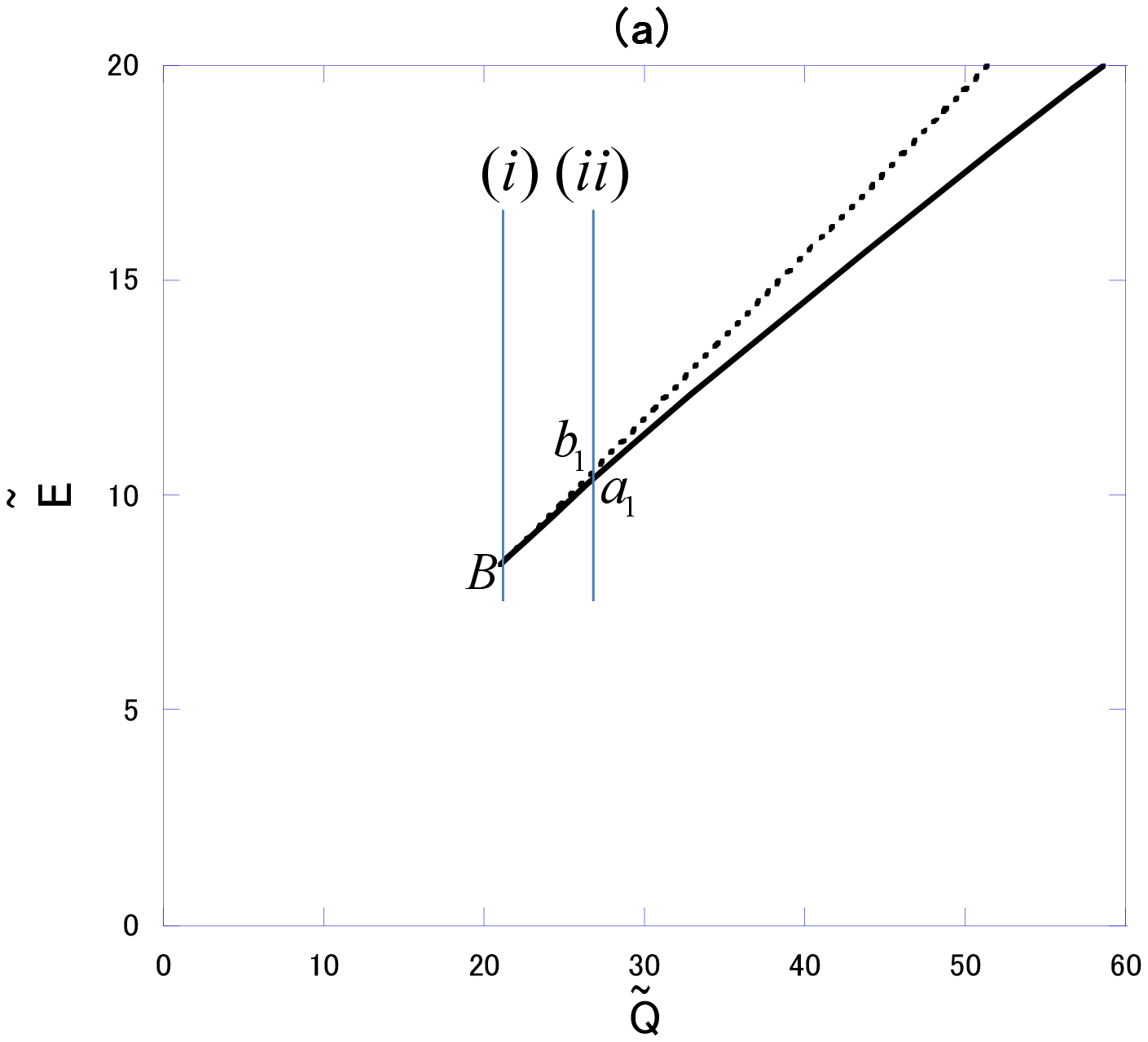,width=3in} \\
\psfig{file=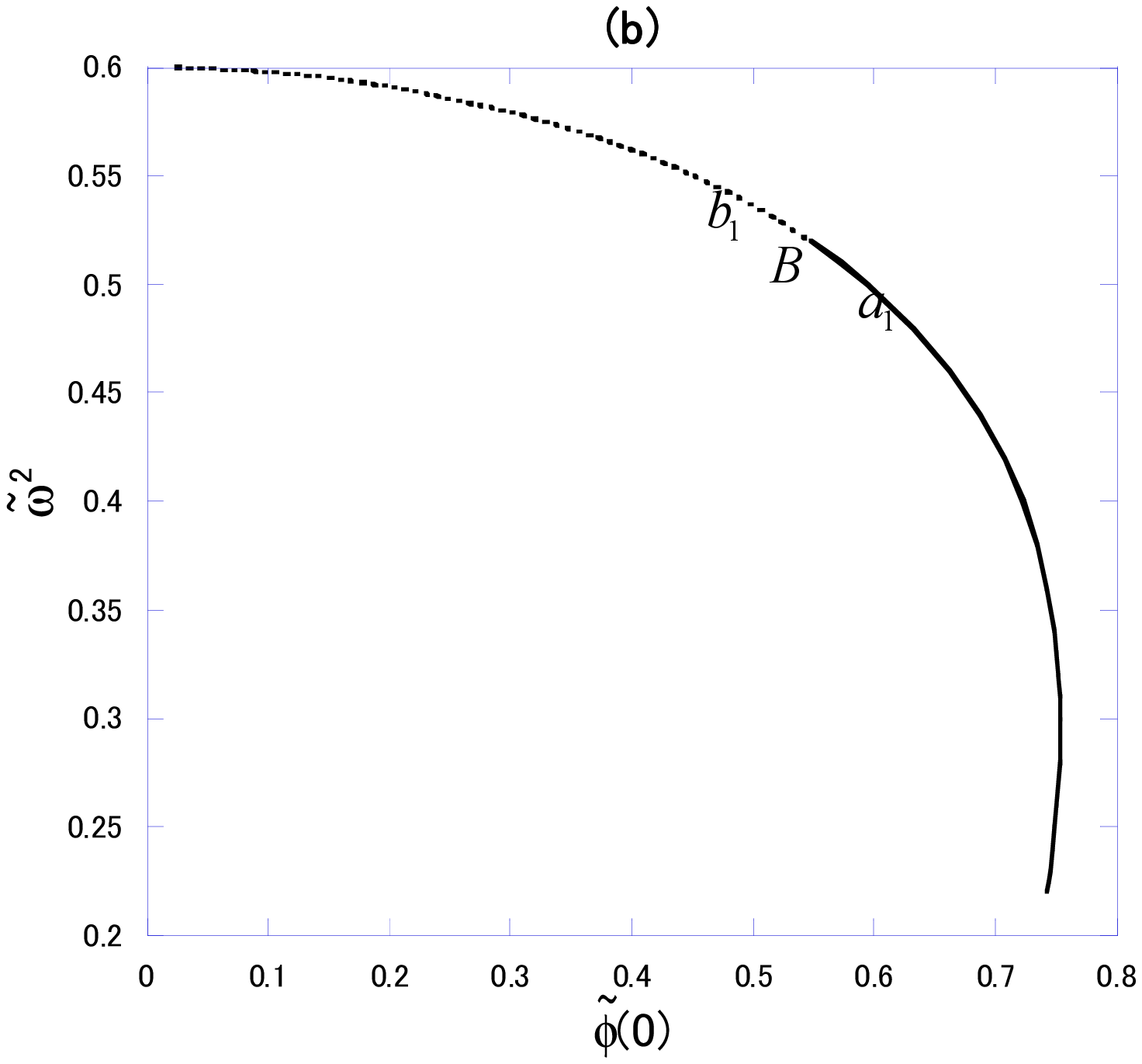,width=3in} \\
\psfig{file=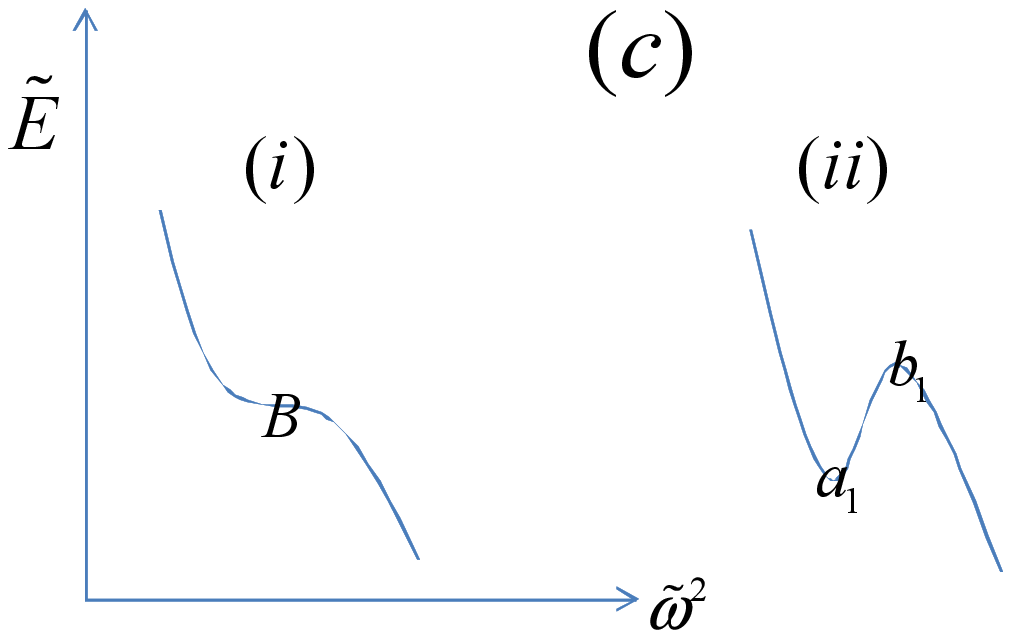,width=3in}
\caption{Stability interpretation via catastrophe theory for $\tilde{m}^2 =0.6$ for the flat case. 
\label{m06catasflat} }
\end{figure}
\begin{figure}[htbp]
\psfig{file=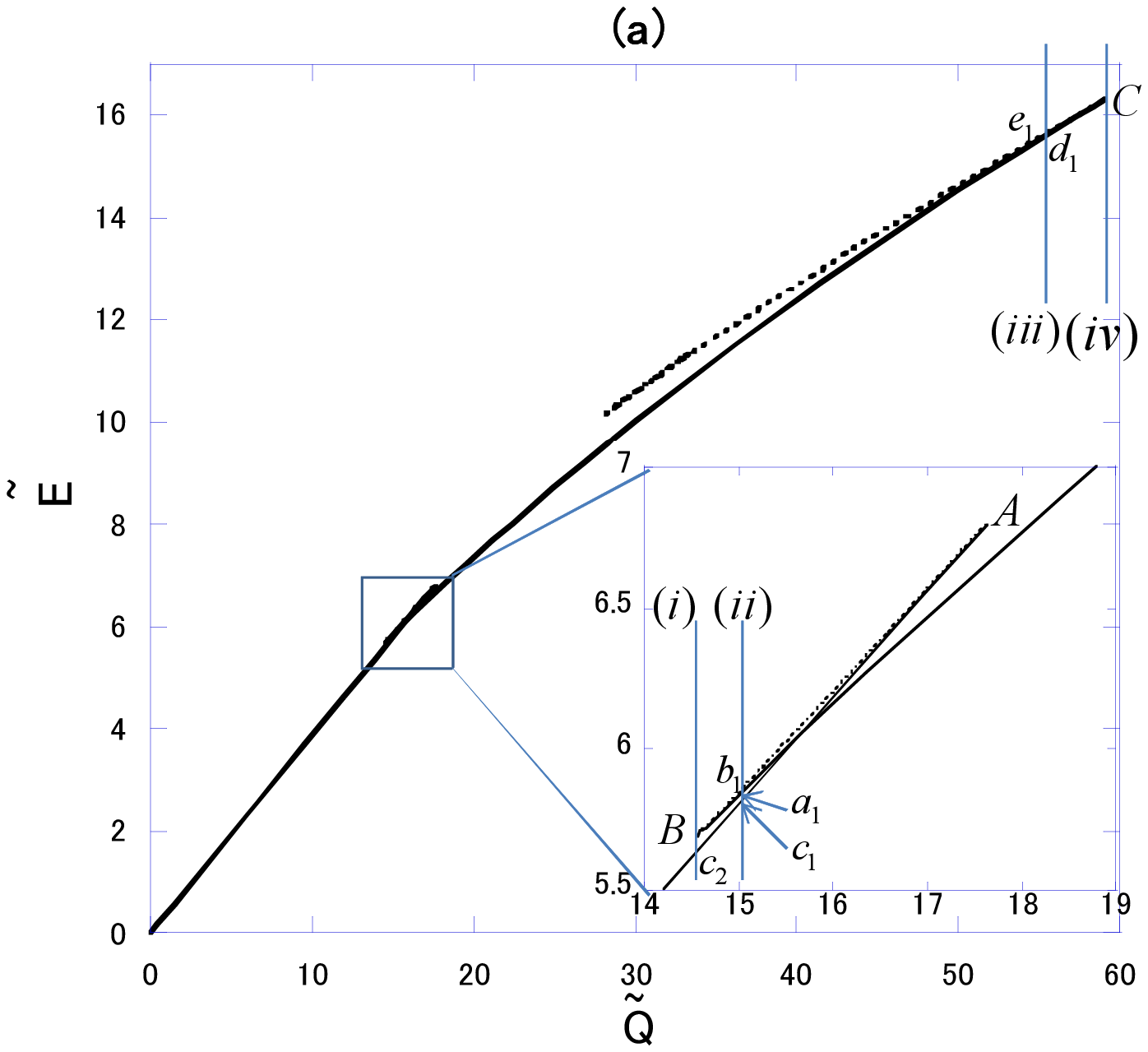,width=3in} \\
\psfig{file=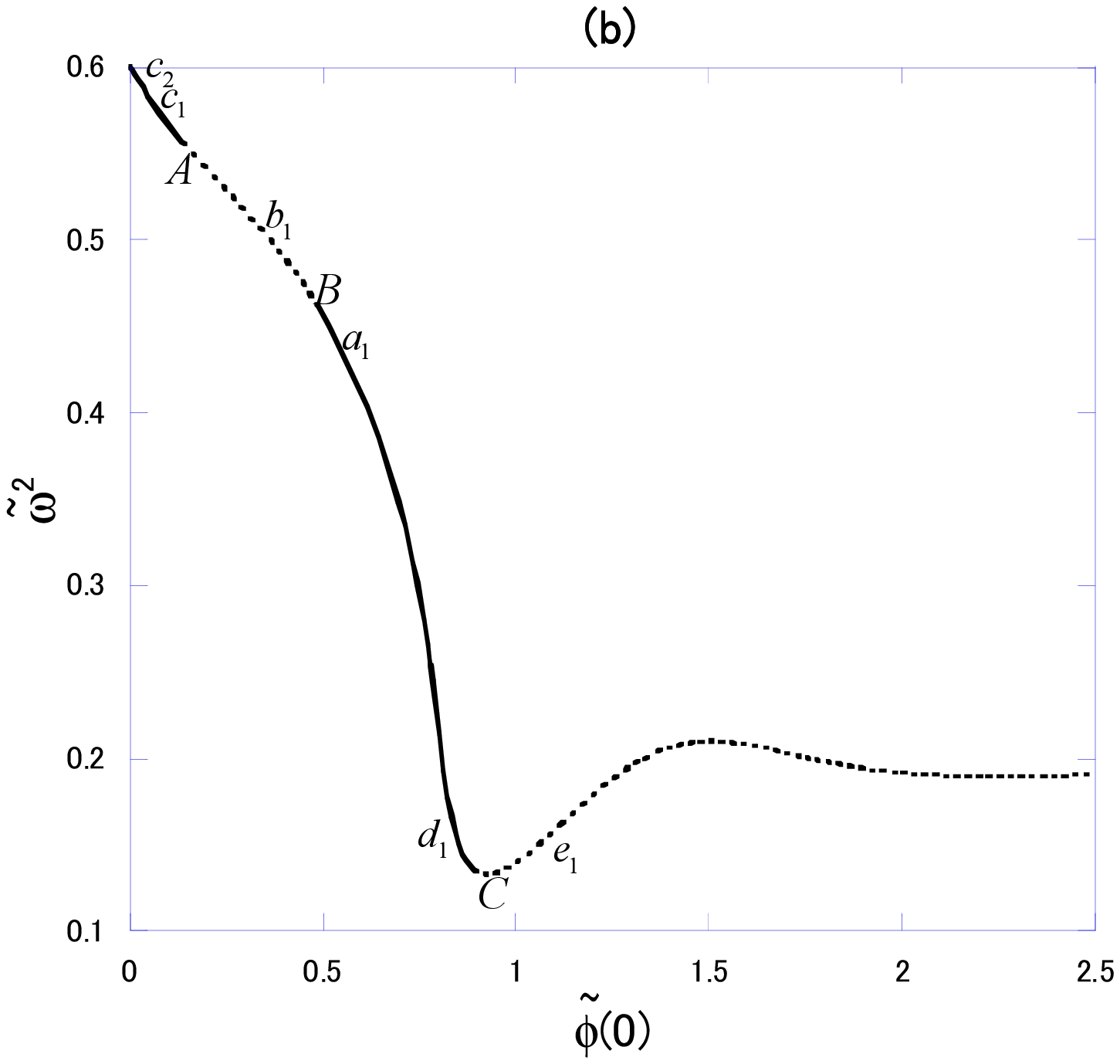,width=3in} \\
\psfig{file=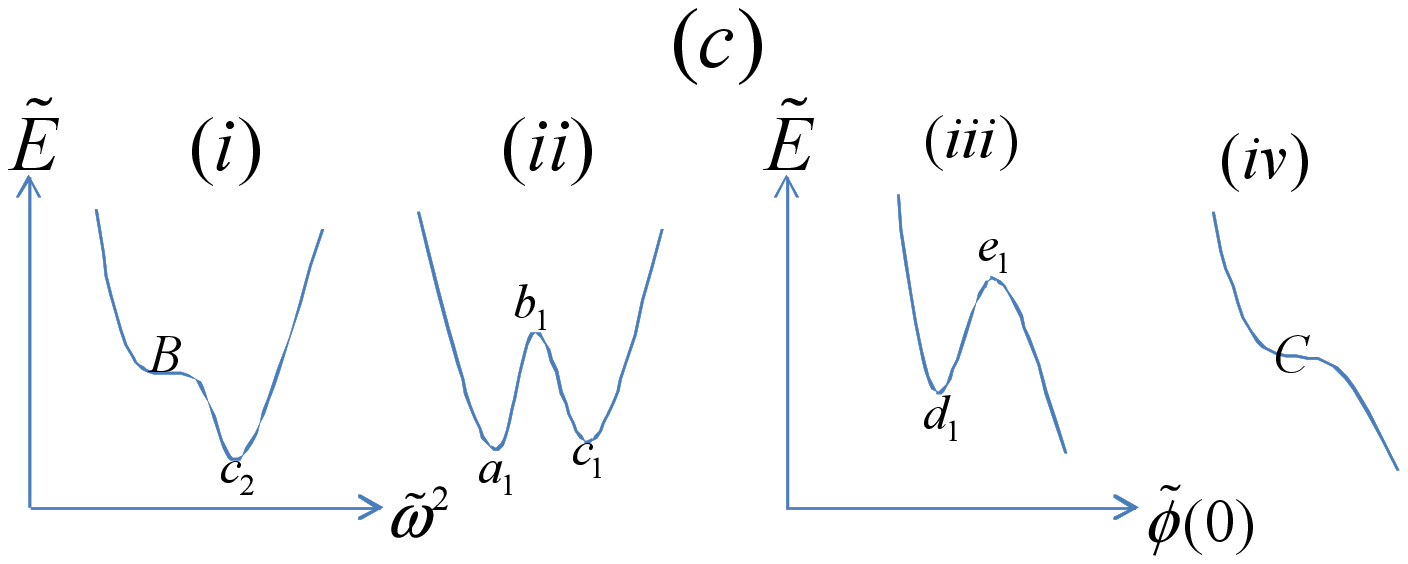,width=3in}
\caption{Stability interpretation via catastrophe theory for $\tilde{m}^2 =0.6$ for $\kappa =0.03$. 
\label{m06catask003} }
\end{figure}
\begin{figure}
\psfig{file=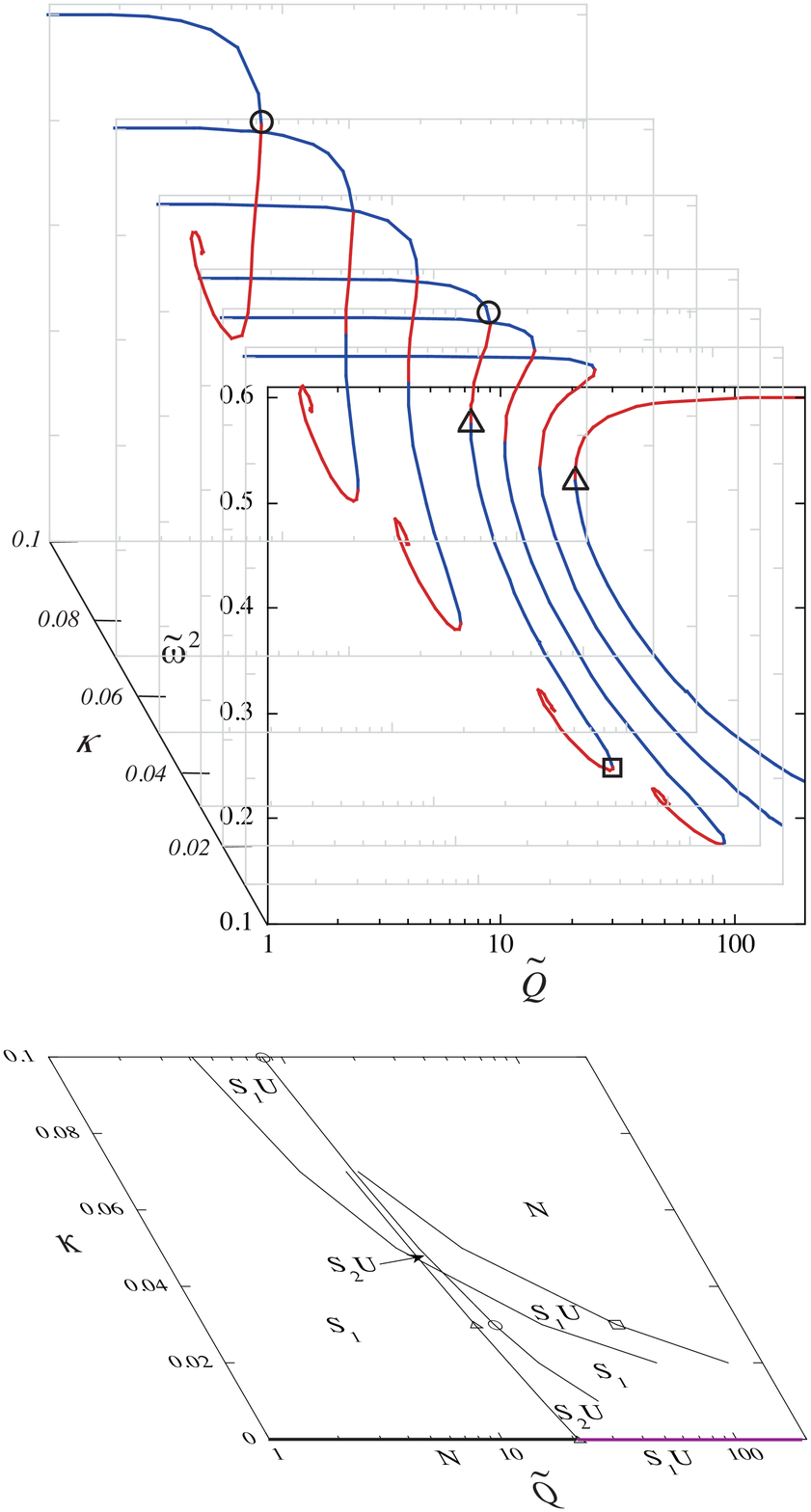,width=3in}  
\caption{\label{m6}
Structures of the {\it equilibrium spaces},
$M=\{(\to^2,\kappa,\tQ)\}$, and their catastrophe map, $\chi(M)$, 
into the {\it control planes}, $C=\{(\kappa ,\tQ)\}$, for $\tilde{m}^2 =0.6$.
Blue lines and red lines in $M$ 
represent stable and unstable solutions, respectively. 
In the regions denoted by S$_{1}$, S$_{i}$U ($i=1,2$) and N on $C$,
there are one stable solution, $i$ stable solution(s) and one or more unstable solution(s),
and no equilibrium solution, respectively, for fixed $(\kappa,\tQ)$.}
\end{figure}

\subsection{Gravitating Q-balls for $\tilde{m}^2 \geq 0.5$}

In this subsection we fix $\tilde{m}^2 =0.6$.
First, we present typical solutions in Figs.~\ref{m06thin} and \ref{m06thick}:
we choose $\kappa =0$, $0.03$, and $0.1$ and show
the metric $A(r)$ in (a) and the scalar field amplitude $\tilde{\phi}(r)$ in (b).
In Figs.~\ref{m06thin} we put  $\tilde{\omega}^{2}\simeq 0.34$.
We find that as $\kappa$ becomes larger, or gravity is stronger, $|A(r)^{2}-1|$ becomes up to order one, and the Q-ball size becomes smaller by self-gravity.
As we shall discuss below, the solutions with $\kappa =0$, $0.03$ in Figs.~\ref{m06thin} are stable, while the solution with $\kappa=0.1$ is unstable.
That is, strong gravity destabilizes or kills some of the solutions which would be existent and stable without gravity.

Figs. \ref{m06thick} show the solutions with $\tilde{\omega}^{2}\simeq 0.595$.
Because $\epsilon^2=0.05\ll1$, these are thick-wall solutions.
We find an interesting feature in (b):
the difference between $\tp$ with $\kappa=0.03$ and $\kappa=0.1$ are small, but they are quite different from $\tp(r)$ with $\kappa=0$.
This indicates that the configuration of $\tp(r)$ for gravitating Q-balls does not approach that for $\kappa=0$ if we take the limit of $\kappa\ra0$.
In the next section we shall discuss the reason for this.

In this way we calculate equilibrium solutions numerically for various $\to^2$ and 
show $\tilde{Q}$-$\tilde{E}$ and $\tilde{Q}$-$\tilde{\omega}^2$ relations in Figs.~\ref{QEml05}.
We can obtain stability of the solutions using catastrophe theory as follows.
\begin{itemize}
\item When there are multiple values of $\tE$ for a given set of the control parameters 
$(\tm^2,\kappa,\tQ)$, by energetics the solution with the lower value of $\tE$ should be stable.
\item Once the stability for a given set of the parameters $(\tm^2,\kappa,\tQ)$ is found, the stability for all sets of parameters which are reached continuously from that 
set without crossing turning points (i.e., stability-change points $\Sigma$) is the same.
\item Stability changes across $\Sigma$.
\item Spiral structure in the $\tQ$-$\to^2$ plane should be considered exceptionally. We interpret that 
all solutions are unstable there.
\end{itemize}
As a result, we can conclude that solid and dashed lines correspond to stable and unstable solutions, respectively. 

To illustrate this energetic or catastrophic argument more clearly, we give a sketch of the potential function $\tE$ near the equilibrium solutions in Figs.~\ref{m06catasflat} and \ref{m06catask003}.
Figs.~\ref{m06catasflat} shows the flat case ($\kappa=0$):
the solid (dotted) line in (a) and the points $B$, $a_{1}$ and $b_{1}$ correspond to those in (b),
where candidates of behavior variables $\tp (0)$ and $\to^{2}$ are shown. (c) is a schematic picture of the potential function $\tE$ in terms of the behavior variable $\tilde{\omega}^2$.
The point B, the cusp in (a), corresponds to the inflection point in (c);
we understand why there is no solution with lower $\tQ$.
The point $a_{1}$ on the solid line and the point $b_{1}$ on the dotted line in (ii) correspond to 
the potential minimum and the maximum in (c), respectively. 

As another example for catastrophic interpretation, we discuss stability of the solutions for $\kappa =0.03$, using Figs.~\ref{m06catask003}.
A complicated structure appears in the enlargement in (a).
In the $Q$-range between $A$ and $B$ there are triple values of $\tE$ for fixed $Q$.
In this case the potential function should be given by (ii) in (c).
This means that two stable solutions coexist for fixed $Q$ in this range.
As a result, we can conclude that there are stable gravitating Q-balls 
which approach  $\tQ\ra0$ in the thick-wall limit ($\tilde{\omega}^{2}\ra 0.6$).
Fig.~\ref{m06catask003} (b) tells us that in the unstable sequence right 
the point $C$ there is no one-to-one correspondence between $\to^2$ and the solutions while there is one-to-one correspondence between $\tp (0)$ and the solutions. 
In this range, therefore, $\tilde{\phi}(0)$ is more appropriate as a behavior variable than $\tilde{\omega}^{2}$ as shown in (iii) and (iv) in (c).

\begin{figure}[htbp]
\psfig{file=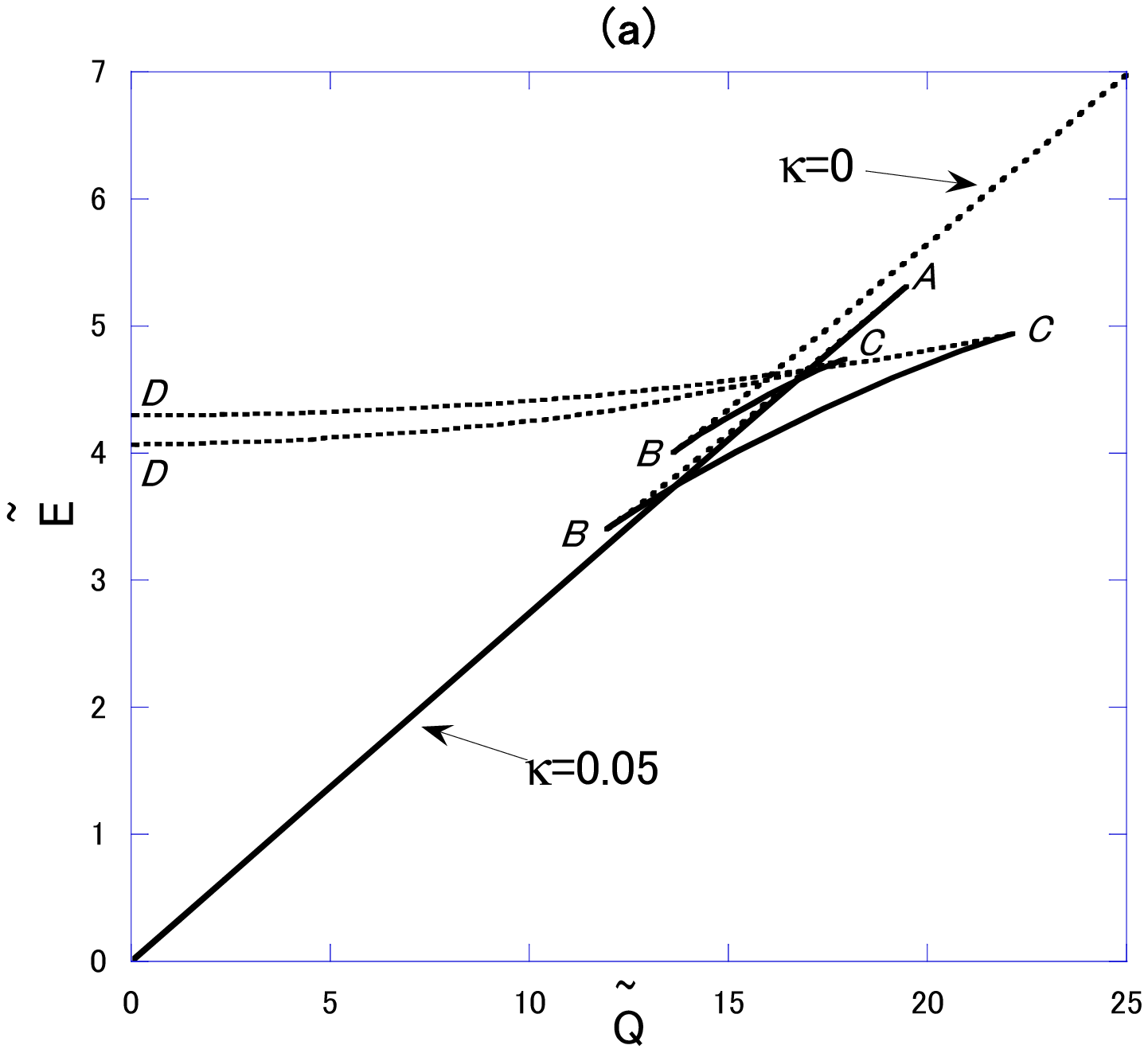,width=3in}  \\
\psfig{file=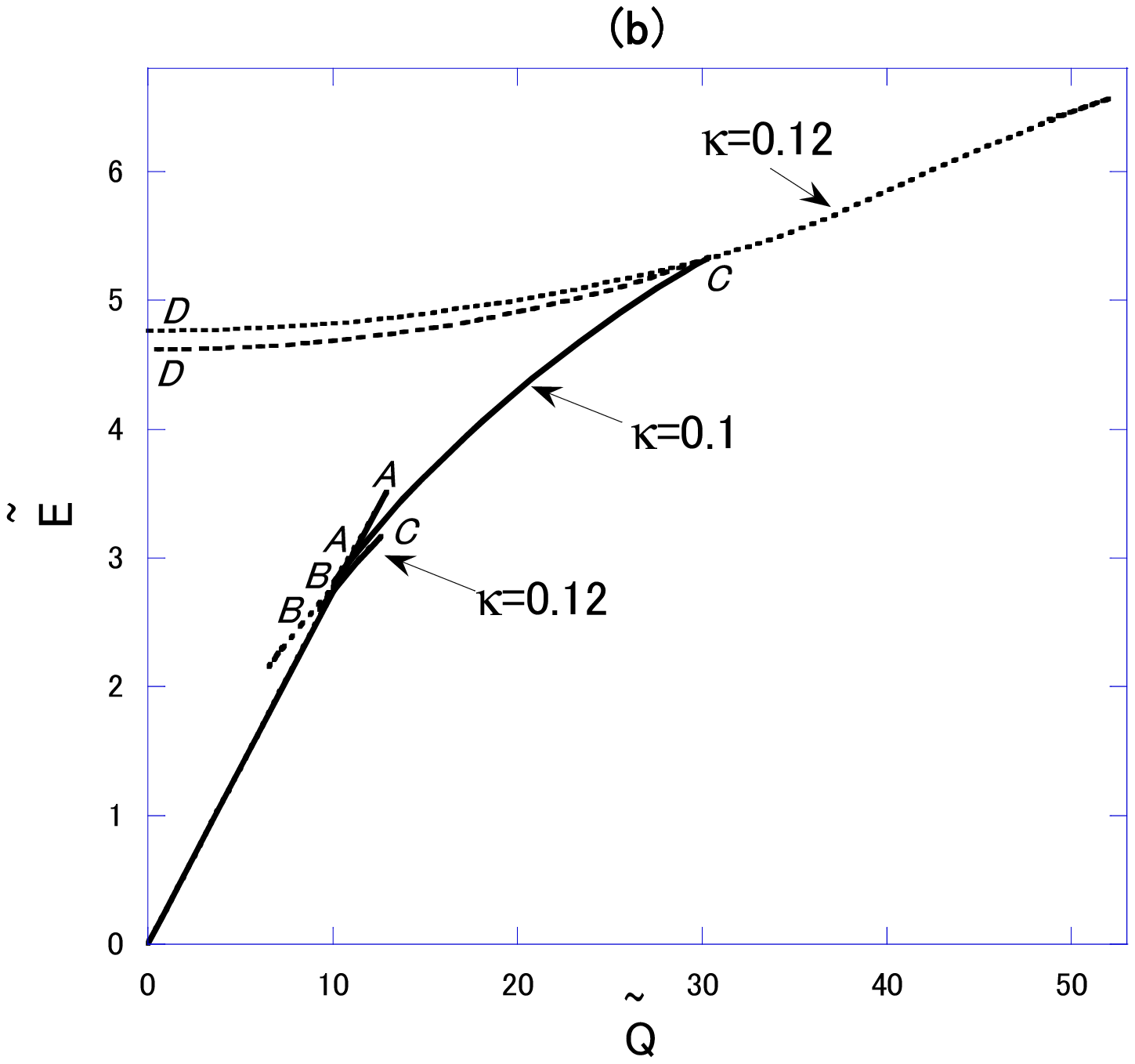,width=3in}
\caption{$\tilde{Q}$-$\tilde{E}$ relations for $\tilde{m}^{2}=0.3$ with $\kappa =0$ and $0.05$ in (a) 
and with $\kappa =0.1$ and $0.12$ in (b), respectively. 
As in the case for $\tilde{m}^{2}=0.6$, gravitating cases have 
sequences from $A$ to the origin, which are considered to be stable, written by solid lines. 
For $\kappa =0.12$, the sequence from $B$-$C$-$D$ separates into two parts. 
Each branch has sequences of cusp 
structures (around $\tilde{Q}\sim 8$ and $48$ in (b)). 
\label{QEmb05}}
\end{figure}

Fig.~\ref{m6} shows the structures of the {\it equilibrium spaces}, 
${\cal M}=\{(\to^2,\kappa,\tQ)\}$, and their catastrophe map, $\chi({\cal M})$, 
into the {\it control planes}, $C=\{(\kappa ,\tQ)\}$, for $\tilde{m}^2 =0.6$. 
In the regions denoted by S$_{1}$, S$_{i}$U ($i=1,2$) 
and N on $C$, there are one stable solution, $i$ stable solution(s) and one or 
more unstable solution(s), and no equilibrium solution, respectively, for fixed $(\kappa,\tQ)$. 
The points $A$, $B$ and $C$ in Figs.~\ref{QEml05} are marked by circles, triangles and squares, respectively.
For example, for $\kappa =0.1$, which is the case shown in Figs.~\ref{QEml05}, 
we can confirm that only a stable solution exists below $\tQ \sim 4$ (the point $D$) which is denoted by S$_{1}$ in Fig.~\ref{m6}. One stable solution and one or more unstable solution(s) exist in the region from $\tQ \sim 4$ to $\tQ \sim 8$ (the point $A$) which is denoted by S$_{1}$U. 

Main characteristics of the equilibrium solutions in Figs.~\ref{QEml05} and \ref{m6} are summarized as follows.
\begin{itemize}
\item If $\kappa=0$, there is a minimum charge, $Q_{\rm min}$, denoted by $B$ on the $\kappa=0$ line in Fig.~\ref{QEml05}. The equilibrium solutions in the thick-wall limit $\epsilon^2\ra0$ are unstable, as indicated by the dotted lines.
 \item If $\kappa\ne0$, no matter how small $\kappa$ is, the equilibrium solutions in the thick-wall limit $\epsilon^2\ra0$ are stable and $\tQ\ra0$.  These stable solutions correspond to the solid lines from $\tQ=0$ to $A$  in Fig.~\ref{QEml05}. We can interpret that gravity saves thick-wall Q-balls. 
\item If $\kappa\ne0$, the maximum charge, $Q_{\rm max}$, emerges in the thin-wall range. This extreme solution is denoted by $C$ on the $\kappa=0.03$ line in Fig.~\ref{QEml05}. That is, gravity kills Q-balls with large charge.
\item If $\kappa\ne0$, spiral trajectories appear in the $\tQ$-$\to^2$ plane.
\item If $\kappa\simeq0.1$, the two extremal solutions $B$ and $C$ merge and disappear, and accordingly the stability sequence between them disappear, too. 
\end{itemize}
The second result is remarkable. 
In Sec.~IV, we investigate why these discontinuous changes occur at $\kappa =0$. 

\begin{figure}
\psfig{file=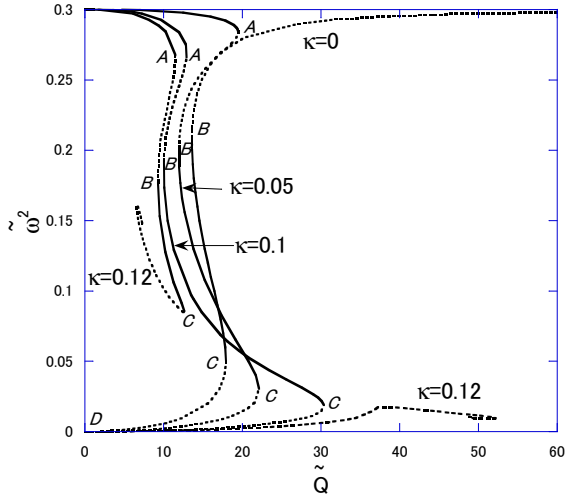,width=3in}  
\caption{\label{Qw2mb05}
$\tQ$-$\tilde{\omega}^{2}$ relation for the same solutions as in Figs.~\ref{QEmb05}. 
The points from $A$ to $D$ and the dotted (solid) lines correspond to those 
in Figs.~\ref{QEmb05}. From this correspondence, we find that flat Q-balls approach 
$\tilde{Q}\ra \infty$ in the thick-wall limit $\tilde{\omega}^{2}\ra 0.3$ while gravitating 
Q-balls approach $\tilde{Q}\ra 0$ in this limit as in the case for $\tilde{m}^{2}=0.6$. 
}
\end{figure}
\begin{figure}[htbp]
\psfig{file=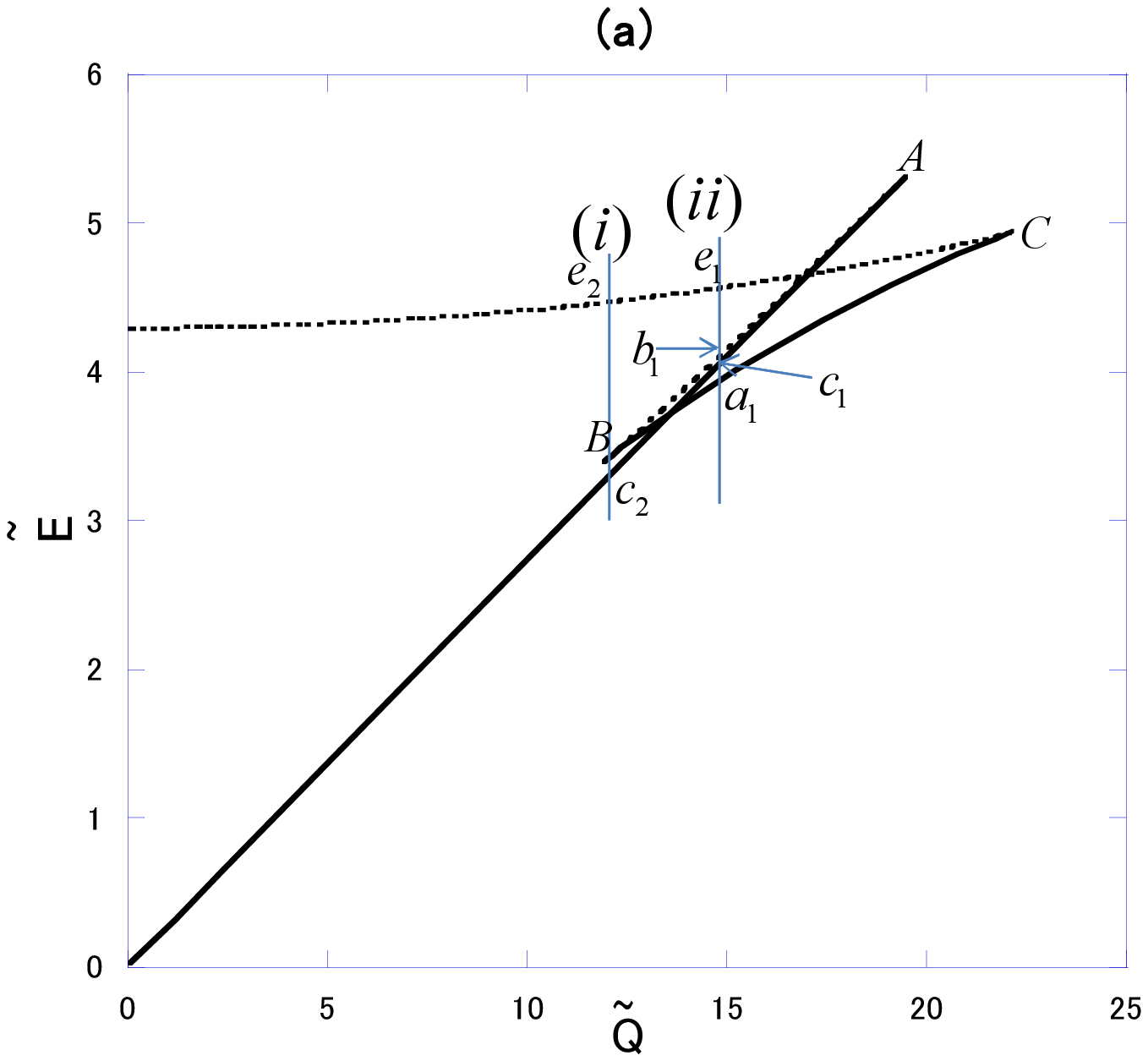,width=2.5in} \\
\psfig{file=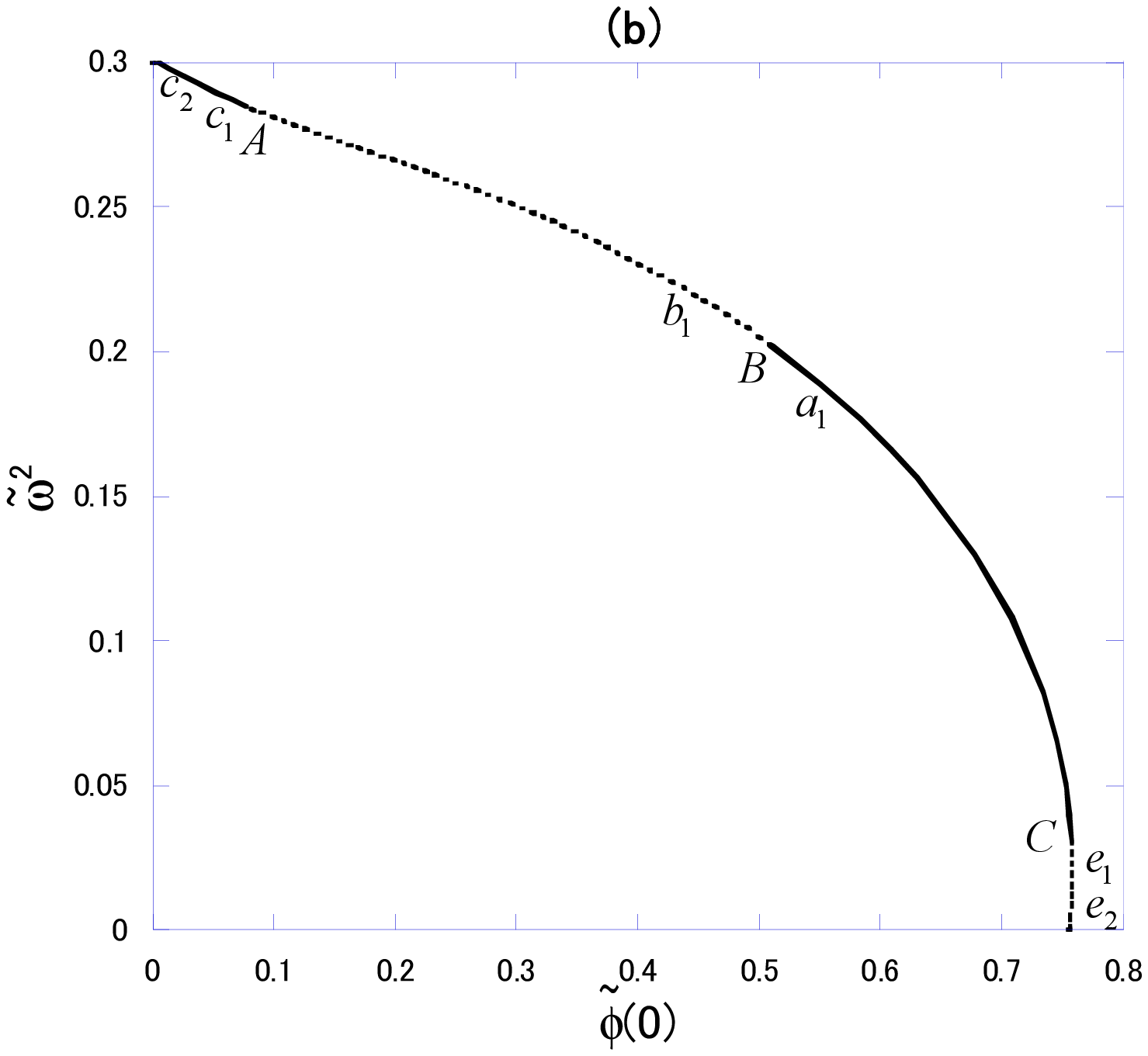,width=2.5in} \\
\psfig{file=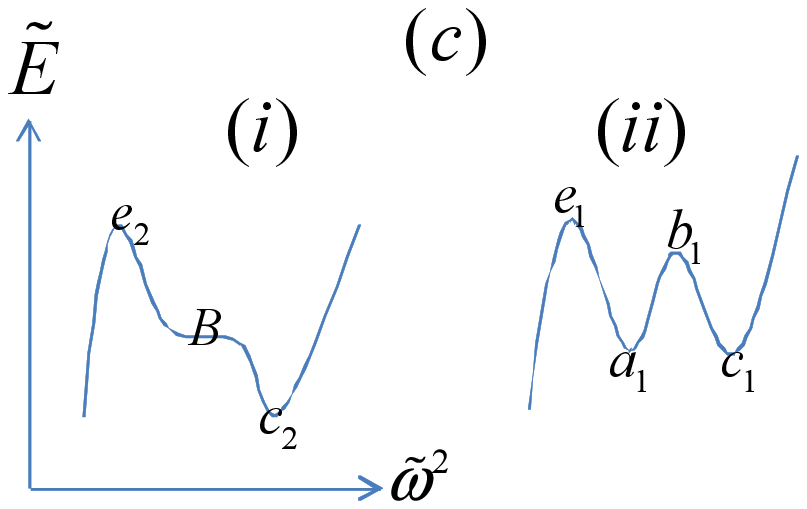,width=2.5in}
\caption{Stability interpretation via catastrophe theory for $\tilde{m}^2 =0.3$ and $\kappa =0.05$. 
\label{m03catask005} }
\end{figure}
\begin{figure}
\psfig{file=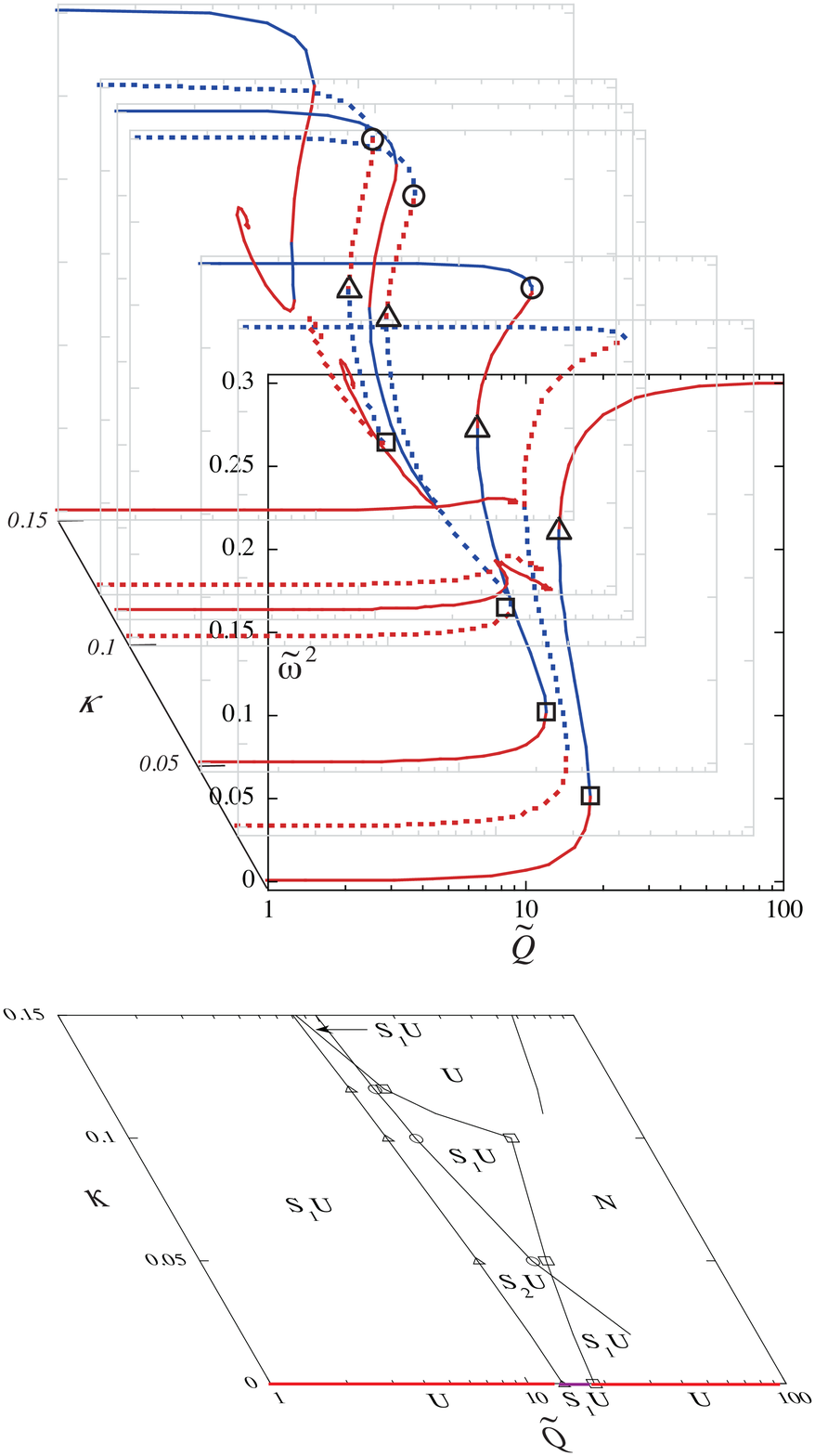,width=3in}  
\caption{\label{m3}
Structures of the {\it equilibrium spaces},
$M=\{(\to^2,\kappa,\tQ)\}$, and their catastrophe map, $\chi(M)$, 
into the {\it control planes}, $C=\{(\kappa ,\tQ)\}$, for $\tilde{m}^2 =0.3$.
Blue lines and red lines in $M$ 
represent stable and unstable solutions, respectively. 
In the regions denoted by S$_{i}$U ($i=1,2$) and N on $C$,
there are $i$ stable solution(s) and one or more unstable solution(s),
and no equilibrium solution, respectively, for fixed $(\kappa,\tQ)$.}
\end{figure}

\subsection{Gravitating Q-balls for $\tilde{m}^2<0.5$}

In this subsection we fix $\tm^2=0.3$.
We show $\tilde{Q}$-$\tilde{E}$ relations in Figs.~\ref{QEmb05} and $\tQ$-$\to^2$ relations in Fig.~\ref{Qw2mb05}.
In the same method as in Sec. III A, we can determine stability of the equilibrium solutions: solid lines and dashed lines correspond to stable and unstable solutions, respectively.

In the case of $\kappa=0$, for example, there are two cusp structures as shown 
in Figs.~\ref{QEmb05} (a). 
Only solutions in the narrow range between $B$ and $C$ are stable.
As another example, for $\kappa=0.05$ we illustrate catastrophic interpretation in Figs.~\ref{m03catask005}.
In the $Q$-range between $A$ and $B$ there are quadruple values of $\tE$ for fixed $\tQ$.
In this case the potential function is given by (ii) in (c).
As in the case of $\tm^2=0.6$ and $\kappa=0.03$ in Fig.~\ref{m06catask003},  two stable solutions coexist for fixed $Q$ in this range, and there are stable 
gravitating Q-balls which approach $\tQ\ra0$ in the thick-wall limit 
($\tilde{\omega}^{2}\ra 0.3$).

Figure \ref{m3} shows the structures of the {\it equilibrium spaces}, 
${\cal M}=\{(\to^2,\kappa,\tQ)\}$, and their catastrophe map, $\chi({\cal M})$, 
into the {\it control planes}, $C=\{(\kappa ,\tQ)\}$, for $\tilde{m}^2 =0.3$. 
$\chi({\cal M})$ shows that in the regions denoted by S$_{i}$U ($i=1,2$) 
and N on $C$, there are $i$ stable solution(s) and one or 
more unstable solution(s), and no equilibrium solution, respectively, for fixed $(\kappa,\tQ)$. 
The points $A$, $B$ and $C$ in Figs.~\ref{QEmb05} are marked by circles, triangles and squares, respectively. 
For example, for $\kappa =0.05$, if we fix $\tQ$ between $\simeq 12$ (the point $B$) and $\simeq 18$ (the point $A$), 
there are two stable solutions and one (or more) unstable solution(s). 

Main characteristics of the equilibrium solutions in Figs. \ref{QEmb05}, \ref{Qw2mb05} 
and \ref{m3} are summarized as follows.
\begin{itemize}
\item If $\kappa=0$, there is a maximum charge for stable solutions, $Q_{\rm max}$, denoted by $C$, as well as a minimum charge, $Q_{\rm min}$, denoted by $B$, on the $\kappa=0$ line in Figs.~\ref{QEmb05}, \ref{Qw2mb05}.
The equilibrium solutions in the thick-wall limit $\epsilon^2\ra0$ are unstable, as indicated by the dotted lines.
\item If $\kappa\ne0$, no matter how small $\kappa$ is, the equilibrium solutions in the thick-wall limit $\epsilon^2\ra0$ are stable and $\tQ\ra0$.  These stable solutions correspond to the solid lines from $\tQ=0$ to $A$  in Figs.~\ref{QEmb05}. 
We can interpret that gravity saves thick-wall Q-balls. 
\item As $\kappa$ increases, the maximum charge, $Q_{\rm max}$, increases until $\kappa=\kappa_{\rm crit}\simeq0.1$.
\item  The solution sequence for fixed $\kappa$ splits into two when $\kappa=\kappa_{\rm crit}$.
In each sequence spiral trajectories appear in the $\tQ$-$\to^2$ plane.
\end{itemize}

\section{Thick-wall limit}


It is not surprising that properties of gravitating Q-balls change gradually as $\kappa$ increases.
It seems strange, however, properties of gravitating Q-balls in the limit of $\kappa\ra0$ differs completely from that of flat Q-balls ($\kappa=0$), as show in Figs.~\ref{m6} and \ref{m3}.
Here we discuss the reason for this.

We consider the case of weak gravity $(\kappa\ll1)$ and thick-wall $(\epsilon^2\ll1)$.
Since the gravity is weak, we can express the metric functions as
\beq\label{weak}
\alpha^{2}=1+h(r), ~~A^2=1+f(r),~~
(h\ll1, ~~ f\ll1).
\eeq
Up to first order in $h$ and $f$, we can rewrite the scalar field equation (\ref{Box}) as
\beq\label{rsfe3dash}
\tp^{\prime\prime}+\left({2\over\tr}+{h'\over2}-{f'\over 2}\right)\tpp
=(1+f)[(\epsilon^2 +h\to^2 )\tp^2-4\tp^3 +6\tp^5]\ .
\eeq
If we fix $\epsilon^2>0$ and take the limit of $\kappa\ra0$ (i.e., $h,~f\ra0)$, Eq.(\ref{rsfe3dash}) 
reduces to the field equation in flat spacetime, (\ref{rsfeflat}).
However, if we take the limit of $\epsilon^2\ra0$ as well, the situation becomes complicated.
For any small $\kappa$, if we take so small $\epsilon^2$ that $\epsilon^2\ll h\to^2$,
which means $\epsilon^2\ll \kappa\to^4$ as we shall show below, 
the first order term in $h$,  $h\to^2 \tp^2$, dominates the zeroth order term, $\epsilon^2\tp^2$. 
That is, in the thick-wall limit of $\epsilon^2\ra0$, the scalar field equation with infinitesimally small $\kappa$ can be different from that with $\kappa=0$.

The above argument is based on the hypothesis that $h$ does not approach zero as fast as $\epsilon^2$ when we take the limit of $\epsilon\ra0$.
Otherwise,  the inequality $\epsilon^2<h\to^2$ would be wrong.
To complete this argument, we shall estimate the order of magnitude of $h$ by assuming
\beq\label{h-estimate0}
\tp (\tr )\sim \epsilon\ll1
~~ {\rm for }~~ \tr <{1\over\epsilon},
\eeq
which is valid in flat spacetime.
From the Einstein equations, we find
\beq\label{h-estimate}
-G^t_t+G^i_i:=\left({\tr^2 \alpha '\over A}\right)'
=8\pi\kappa \tr^2 A\alpha \left(\frac{\to^2 \tp^2}{\alpha^2} -V\right)\ ,
\eeq
where $i$ runs spacial components. 
If we take the weak field approximation (\ref{weak}) and the thick-wall approximation $\epsilon^2\ll1$, we obtain
\beq\label{h-estimate2}
(\tr^2 h ')'\simeq 8\pi\kappa \tr^2\tm^2 \tp^2 \ .
\eeq
With the boundary condition $h'(0)=0$ and the approximation (\ref{h-estimate0}), we can integrate (\ref{h-estimate2}) as 
\beq\label{h-estimate3-1}
h'\simeq \frac83\pi\kappa \tm^2 \epsilon^2 \tr,
~~ {\rm for }~~ \tr <{1\over\epsilon}.
\eeq
With the boundary condition $h(\infty)=0$ and the approximation (\ref{h-estimate0}), we can integrate (\ref{h-estimate3-1}) as
\beq\label{h-estimate3}
h\simeq \frac43\pi\kappa\tm^2 (\epsilon^2\tr^2-1),
~~ {\rm for }~~ \tr <{1\over\epsilon}.
\eeq
This means $h\sim\kappa\tm^2$, which is independent of $\epsilon$.
Therefore, we can conclude $\epsilon^2\ll h\to^2$ in (\ref{rsfe3dash}) in the thick-wall limit 
with fixed $\kappa$.

If the assumption (\ref{h-estimate0}) is not valid, the configuration of $\tp(\tr)$ is quite different from that for flat spacetime. This case also means that gravitating Q-balls 
are completely different from those in flat case even if the gravity is very weak.

From the above argument, we can understand why gravity saves thick-wall Q-balls. 
This is not surprising because similar phenomenon occurs in the case of boson stars:
boson stars with $V_{\rm BS}$ do not exist in flat case while 
they exist even if the gravity is very weak.

\section{Conclusion and discussion}

We have analyzed stability of gravitating Q-balls with $V_{4}$ via catastrophe theory.
Our results are summarized as follows.

Although our original concern was massive Q-balls with astronomical size, we have found an unexpected result that the weak gravity changes properties of thick-wall Q-balls.
In flat spacetime Q-balls in the thick-wall limit are unstable and there is a minimum charge $Q_{{\rm min}}$, where Q-balls with $Q<Q_{{\rm min}}$ are nonexistent.
If we take self-gravity into account, on the other hand, there exist 
stable Q-balls with arbitrarily small charge, no matter how weak gravity is. 
That is, gravity saves Q-balls with small charge. 

This result indicates that gravitational effects may be important for other models, such as Q-balls in supersymmetric extensions of the Standard Model.
For example, gravity may allow for a new branch of solutions in some parameter range where equilibrium solutions are nonexistent in the absence of gravity.

We have also shown how stability of Q-balls changes as gravity becomes strong.
For example,  if $m^2\ge0$, the maximum charge, $Q_{{\rm max}}$, decreases as gravity becomes strong, while there is no maximum charge in flat spacetime.
That is, gravity kills thin-wall Q-balls with large charge.

In the case of strong gravity, only Q-balls with small charge exist, and instability solutions make spiral trajectories in the $\tQ$-$\to^2$ plane.
These properties are common to Q-balls with $V_{3}$ potential and boson stars with $V_{\rm BS}$.
While Q-balls and boson stars have been studied separately so far, our result suggest that there is universal nature of gravity, which may be important to discuss Q-balls with astronomical size or boson stars.

\acknowledgements
We would like to thank Kei-ichi Maeda for useful discussion and for 
continuous encouragement. The numerical calculations were carried out 
on SX8 at  YITP in Kyoto University.
This work was supported by MEXT Grant-in-Aid for Scientific Research on Innovative Areas No.\ 22111502.



\begin{thebibliography}{99}
\bibitem{foot1}
This model has also been considered in \cite{KKL05}. They fixed the parameter 
$\tm^2 >0.5$ and focused on the rotating case.
\end{thebibliography}
\end{document}